    \documentstyle[epsf]{article} \catcode`\@=11
    \def\section{\@startsection{section}{1}{\z@}%
    {-3.5ex plus -1ex minus -.5ex}{1.5ex plus.3ex}{\bf }}
    \def\subsection{\@startsection{subsection}{1}{\z@}%
    {-3.5ex plus-1ex minus-.5ex}{1.5ex plus.3ex}{\bf }} 
    
\begin{document}
    \hfill\parbox{4.77cm}{\Large\centering Annalen\\der
    Physik\\[-.2\baselineskip] {\small \underline{\copyright\ Johann
    Ambrosius Barth 1998}}} \vspace{.75cm}\newline{\Large\bf
 Lax pair formulation for a small-polaron chain with integrable boundaries 
    }\vspace{.4cm}\newline{\bf   
 Xi-Wen Guan,$^{1,2}$ Uwe Grimm$^{1}$ and Rudolf A.\ R{\"o}mer$^{1}$ 
    }\vspace{.4cm}\newline\small
$^{1}$Institut f{\"u}r Physik, Technische Universit{\"a}t, D-09107
  Chemnitz, Germany\\
 $^{2}$Department of Physics, Jilin University, Changchun 130023, China\\
    \vspace{.2cm}\newline 
Received 1 April 1998, revised version 18 September 1998, accepted 6
October 1998
    \vspace{.4cm}\newline\begin{minipage}[h]{\textwidth}\baselineskip=10pt
    {\bf  Abstract.}
Using a fermionic version of the Lax pair formulation, we construct an
integrable small-polaron model with general open boundary conditions. The
Lax pair and the boundary supermatrices $K_{\pm}$ for the model are
obtained. This provides a direct proof of the integrability of the model.
    \end{minipage}\vspace{.4cm} \newline {\bf  Keywords:}
Correlated electrons, Yang-Baxter equation, 
integrable boundary conditions, Lax pair formulation
    \newline\vspace{.2cm} \normalsize

\section{Introduction}
\label{sec1}

Recently, there has been renewed interest in integrable boundary
systems \cite{ref1,ref2} due to their connection to the Kondo problem
\cite{kondo} and boundary conformal field theory \cite{BCFT} in
low-dimensional quantum many-body systems. Sklyanin \cite{ref3}
proposed a systematic approach to construct and solve integrable
quantum spin systems with open boundary conditions (BC).  Central to
his method is the introduction of the so-called reflection equations
(RE), which are the boundary analogues of the Yang-Baxter equations
(YBE) \cite{baxter}. For integrable fermion systems, in particular
models of strongly correlated electrons, graded versions of the YBE
and the RE \cite{ref4} have been applied. However, solutions to the
graded RE have so far been found only for certain models due to the
mathematical complexity of the graded RE. An alternative method to
Sklyanin's approach is the Lax pair formulation for integrable quantum
systems with open BC \cite{ref5}.  In this article, we use the Lax
pair formulation to construct an integrable model of correlated
fermions with general BC.  The Lax pair and the boundary supermatrices
$K_{\pm}$ are given explicitly. 

\section{Fermionic version of the Lax pair formulation}
\label{sec2}

We first recall the Lax pair formulation for integrable fermion models
with general open BC in one dimension \cite{ref5}.  We consider an operator
version of the auxiliary linear problem
\begin{eqnarray}
\Phi_{j+1} & = & L_j(u)\Phi_j,\quad j=1,2,\ldots ,N,\nonumber\\
\frac{d}{dt}\Phi_j & = & M_j(u)\Phi_j,\quad j=2,3,\ldots ,N,\label{2a}
\end{eqnarray}
with boundary equations
\begin{eqnarray}
\frac{d}{dt}\Phi_{N+1} & = & M_+(u)\Phi _{N+1}\nonumber\\
\frac{d}{dt}\Phi_1 & = & M_-(u)\Phi_1.\label{2aa}
\end{eqnarray}
Here, $L_j(u)$, $M_j(u)$, and $M_{\pm}(u)$ are matrices depending on
the spectral parameter $u$; $u$ itself does neither depend on the time
$t$, nor on dynamical variables. Evidently, the consistency conditions
for Eqs.~(\ref{2a}) and (\ref{2aa}) yield the Lax equations
\begin{equation}
\frac{d}{dt}L_j(u)=M_{j+1}(u)L_j(u)-L_j(u)M_j(u),
\quad j=2,3,\ldots ,N-1,\label{2b}
\end{equation}
and 
\begin{eqnarray}
\frac{d}{dt}L_N(u)&=&M_+(u)L_N(u)-L_N(u)M_N(u),\nonumber\\
\frac{d}{dt}L_1(u)&=&M_2(u)L_1(u)-L_1(u)M_-(u).\label{2bb}
\end{eqnarray}
Let $T(u)=L_N(u)\cdots L_2(u) L_1(u)$ be the usual monodromy matrix
\cite{baxter} of the system and $K_{-}(u)$, $K_{+}(u)$ the supermatrices for
the left and the right boundary, respectively. The transfer matrix 
$\tau(u)$ is defined as the supertrace on the auxiliary space $V_{0}$ as
\begin{equation}
\tau(u)={\rm str}_0\left[K_+(u)T(u)K_-(u)T^{-1}(-u)\right].\label{t} 
\end{equation}
{}From the Lax equations (\ref{2b}) and (\ref{2bb}), it follows that
the transfer matrix $\tau(u)$ does not depend on time $t$, provided the
boundary matrices satisfy the conditions
\begin{eqnarray}
K_-(u)M_-(-u) & = & M_-(u)K_-(u) \nonumber\\
{\rm str}_0\left[K_+(u)M_+(u)\tilde{T}(u)\right] & = & 
{\rm str}_0\left[K_+(u)\tilde{T}(u)M_+(-u)\right],\label{2cd}
\end{eqnarray}
where $\tilde{T}(u)=T(u)K_{-}(u)T^{-1}(-u)$.  This implies that the
system possesses an infinite number of independent conserved
quantities, and thus is completely integrable.

\section{The small-polaron model with general open BC}
\label{sec3}

We consider the small-polaron model \cite{ref6}, which describes the
motion of an additional electron in a polar crystal with external
field parallel to the transverse direction. The Hamiltonian reads
\begin{eqnarray}
H & = & -
J\sum_{j=2}^{N}(a_j^{\dagger}a_{j-1}^{}+a_{j-1}^{\dagger}a_j^{}) + 
V\sum_{j=2}^{N}n_{j}^{}n_{j-1}^{} +
W\sum_{j=1}^{N}n_{j}^{} 
\nonumber\\ 
& & \mbox{}+p_{+}n_{N} + \alpha_{+} a_{N}^{\dagger} + \beta_{+} a_{N}^{} +
p_{-}n_{1}^{}+\alpha_{-}a_{1}^{\dagger} +\beta_{-}a_{1}^{},
\label{2f}
\end{eqnarray}
where in the boundary terms $p_{\pm}$, $\alpha_{\pm}$ and
$\beta_{\pm}$ are Grassmann variables, with $p_{\pm}$ even and
$\alpha_{\pm}$, $\beta_{\pm}$ odd. Hermiticity of the Hamiltonian requires $\alpha _{\pm }^{\dagger }=\beta _{\pm }$ and $p_{\pm }^{\dagger }=p_{\pm }$
The fermionic creation and
annihilation operators $a_{j}^{\dagger}$ and $a_{j}^{}$ satisfy the usual
anticommutation relations, and $n_j^{}=a_{j}^{\dagger}a_{j}^{}$.
We parametrize the coupling parameters $J$, $V$, and $W$ as
\begin{eqnarray}
J & = & 1, \nonumber\\
V & = & -2\cos(2\eta),\nonumber\\
W & = & 2\sin(2\eta)\tan(\omega)+\cos(2\eta),
\end{eqnarray} 
in terms of $\eta$ and $\omega$. The monodromy matrix corresponding to
the Hamiltonian (\ref{2f}) is given in Ref.~\cite{ref7}, and obeys the
graded YBE. This ensures the integrability of the model with periodic
BC. In what follows, we shall use this monodromy matrix and construct
the boundary $K$-matrices as described in Sec.~\ref{sec2} in order
to show the integrability of the Hamiltonian.

\section{Constructing the Lax pair at the boundaries}
\label{Sec4}

The equations of motion $-i\frac{d}{dt}O=[H,O]$ for the fermionic operators
at the left boundary are
\begin{eqnarray}
-i\frac{d}{dt}a_{1}^{\dagger} & = & 
- J a_{2}^{\dagger} + V n_{2}^{}a^{\dagger}_{1}
+ (W+p_{-})a^{\dagger}_{1}+\beta_{-},
\nonumber\\
-i\frac{d}{dt}a_{1}^{} & = & 
Ja^{}_{2}-V n_{2}^{}a_{1}^{} -
(W+p_{-})a_{1}^{}+\alpha_{-},\nonumber\\
-i\frac{d}{dt}n_{1}^{} & = & 
-J(a^{\dagger}_{2} a_{1}^{}+a_{2}^{}a_{1}^{\dagger})-
\alpha_{-}a_{1}^{\dagger}+\beta_{-}a_{1}^{}.
\end{eqnarray}
At the right boundary, they are given by
\begin{eqnarray}
-i\frac{d}{dt}a_N^{\dagger} & = & 
-J a^{\dagger}_{N-1} + V n_{N-1}^{}a^{\dagger}_N+
(W+p_{+}) a^{\dagger}_N+\beta_{+},\nonumber\\
-i\frac{d}{dt}a_{N}^{} & = & 
Ja^{}_{N-1} - V n_{N-1}^{} a_{N}^{}-(W+p_{+})a_{N}^{}+\alpha_{+},
\nonumber\\
-i\frac{d}{dt}n_{N}^{} & = & 
J(a^{\dagger}_N a_{N-1}^{}+a_N^{}a_{N-1}^{\dagger})-
\alpha_{+} a_N^{\dagger} + \beta_{+}a_{N}^{}.
\end{eqnarray}
Using the Lax matrices $L_{j}(u)$ and $M_{j}(u)$ for the bulk part as
given in Ref.~\cite{ref7}, together with Eq.~(\ref{2bb}), we can construct
the boundary Lax matrices
\begin{equation}
M_{\pm}(u)=\left(\matrix{A_{\pm}(u) & B_{\pm}(u)\cr
                         C_{\pm}(u) & D_{\pm}(u)}\right).
\end{equation}
After some algebra, we find
\begin{eqnarray}
A_{\pm}(u) & = & 
\frac{1}{s_{+2}s_{-2}}\left[
i\sin^{2}2\eta\: (p_{\pm}+\cos 2\eta)(1-n_{\pm}^{}) +
s_{\pm 2}\left(i s_{\mp 2} \pm 
\xi^{\pm 2}_{+}\sin u\right)\alpha_{\pm}a_{\pm}^{\dagger}
\right.\nonumber\\
& & \mbox{\qquad}\left.
+ s_{\mp 2}\left(is_{\pm 2}\mp \xi^{\mp 2}_ +
  \sin u\right)\beta_{\pm}a_{\pm}^{}\rule[0pt]{0pt}{10pt}\right]
-i\sin 2\eta \tan\omega,\\
D_{\pm }(u) & = &
\frac{1}{s_{+2}s_{-2}}\left[
-i\sin^{2}2\eta\: (p_{\pm }-\cos 2\eta)n_{\pm}^{}+
s_{\mp 2}\left(is_{\pm 2}\mp 
\xi^{\pm 2}_{+}\sin u\right)\alpha_{\pm}a_{\pm}^{\dagger}\right.\nonumber\\
& &\mbox{\qquad}\left.
+ s_{\pm 2}\left(is_{\mp 2}\pm 
\xi^{\mp 2}_+\sin u\right)\beta_{\pm}a_{\pm}^{}\rule[0pt]{0pt}{10pt}\right]
+i\sin 2\eta \tan\omega,\\
B_{\pm}(u) & = &
\frac{i^{\frac{1}{2}\pm \frac{1}{2}}}{s_{+2}s_{-2}}\left[\rule[0pt]{0pt}{10pt}
-\xi^{\pm 1}_{+}\sin u\sin 4\eta\: \alpha_{\pm}n_{\pm}^{}\mp 
i\xi^{\mp 1}_{+}\sin 2\eta\: (p_{\pm}\sin u\pm \sin 2\eta \cos u)a_{\pm}^{}
\right.\nonumber\\
& & \mbox{\qquad}\left.
 +\xi^{\pm 1}_+\sin 2\eta\: s_{\mp 2}\alpha_{\pm}\rule[0pt]{0pt}{10pt}
\right],\\
C_{\pm}(u) & = & 
\frac{i^{\frac{1}{2}\pm\frac{1}{2}}}{s_{+2}s_{-2}}\left[
\mp \xi^{\mp 1}_+\sin u\sin 4\eta\: \beta_{\pm}n_{\pm}^{} -
i\xi^{\pm 1}_{+}\sin 2\eta\: \left(p_{\pm}\sin u\mp 
\sin 2\eta \cos u\right)a_{\pm}^{\dagger}\right.\nonumber\\
& & \mbox{\qquad}\left.
\pm \xi^{\mp 1}_{+}\sin 2\eta\: s_{\pm 2}\beta_{\pm}\rule[0pt]{0pt}{10pt}
\right].
\end{eqnarray}
For convenience, we introduced the abbreviations 
\begin{eqnarray}
& & \makebox[4cm][l]{$s_{\pm n}=\sin(u\pm n\eta ),$}
    \makebox[4cm][l]{$c_{\pm n}=\cos(u\pm n\eta ),$} \nonumber\\
& & \makebox[4cm][l]{$a_{+}^{}=a_{N}^{},$}
    \makebox[4cm][l]{$a_{-}^{}=a_{1}^{},$}
\end{eqnarray}
and 
\begin{equation}
\xi_{\pm}=\xi(\pm u)=\frac{\cos(u\pm\omega)}{\cos u \,\cos\omega}.
\end{equation}

\section{Boundary reflection supermatrices}

We are now ready to compute the boundary $K$-matrices
\begin{equation}
K_{\pm}(u)=\left(\matrix{K_{11}^{\pm}&K_{12}^{\pm}\cr 
                         K_{21}^{\pm}&K_{22}^{\pm}}\right)
\label{2g}
\end{equation}
by using Eq.~(\ref{2cd}). This yields 16 equations for the
four matrix elements $K_{ij}^{\pm}$, only six of which are independent.
After a lengthy calculation, we find
\begin{eqnarray}
K_{11}^{-} & = & 
\xi_{+}\sin(u-\psi_{-})
\left(\xi ^2_-s_{+2}-\xi ^2_+s_{-2}\right),\\
K_{22}^{-}& = & 
-\xi_{-}\sin(u+\psi_{-})
\left(\xi^2_+s_{+2}-\xi ^2_-s_{-2}\right),\\
K_{12}^{-} & = & 
-\frac{\alpha_{-}\sin \psi_{-}\sin u}{i\xi_{+}\xi_{-}\sin^{2}2\eta}
\left(\xi^2_{-}s_{+2}-\xi^2_{+}s_{-2}\right)
\left(\xi^2_{+}s_{+2}-\xi^2_{-}s_{-2}\right),\\
K_{21}^{-} & = & 
-\frac{\beta_{-}\sin\psi_{-}\sin u}{i\sin^{2}2\eta}
\left(\xi^2_{-}s_{+2}-\xi^2_{+}s_{-2}\right)
\left(\xi^2_{+}s_{+2}-\xi^2_{-}s_{-2}\right),\\
K_{11}^{+} & = & 
\xi_{+}\sin(u+2\eta -\psi_{+})
\left(\xi^2_{-}s_{+4}-\xi^2_{+}\sin u\right),\\
K_{22}^{+}& = & 
\xi_{-}\sin(u+2\eta +\psi_{+})
\left(\xi^2_{+}s_{+4}-\xi^2_{-}\sin u\right),\\
K_{12}^{+} & = & 
-\frac{\alpha_{+}\sin\psi_{+}s_{+2}}{i\xi_{+}\xi_{-}\sin^{2}2\eta }
\left(\xi^2_{-}s_{+4}-\xi^2_{+}\sin u\right)
\left(\xi^2_{+}s_{+4}-\xi^2_{-}\sin u\right),\\
K_{21}^{+} & = & 
-\frac{\beta_{+}\sin\psi_{+}s_{+2}}{i\sin^{2}2\eta}
\left(\xi^2_{-}s_{+4}-\xi^2_{+}\sin u\right)
\left(\xi^2_{+}s_{+4}-\xi^2_{-}\sin u\right),
\end{eqnarray}
where we have parametrized $p_{\pm}=\sin 2\eta \cot\psi_{\pm}$. The 
Hamiltonian (\ref{2f}) can be found as usual as an invariant of
the commuting family of transfer matrices $\tau(u)$ (\ref{t}) by taking the
derivative at a special value of the spectral parameter $u$. Namely,
\begin{equation}
-\sin 2\eta\:\left. \frac{d}{du}\tau(u)\right|_{u=0} =
2 H \tau(0) + 
{\rm str}_0\left(\left.\frac{d}{du}K_{+}(u)\right|_{u=0} \right) .
\end{equation}

\section{Conclusion}

We have presented the Lax pair formulation for the 1D small-polaron
model with general boundary conditions, providing a direct
demonstration for the integrability of the model. The Grassmann
variables at the boundary terms play a crucial role in the
construction of the boundary $K$-supermatrices. They may be
interpreted in physical terms as sources and sinks for injecting
additional particles into the system. In particular, we can choose as
a special representation at the left end
$\alpha_{-}=c^{(\alpha)}_{-}a_{0}^{}$,
$\beta_{-}=c^{(\beta)}_{-}a_{0}^{\dagger}$, and $p_{-}=c^{(p)}_{-}$,
and at the right end $\alpha_{+}=c^{(\alpha)}_{+}a_{N+1}^{}$,
$\beta_{+}=c^{(\beta)}_{+}a_{N+1}^{\dagger}$, and $p_{+}=c^{(p)}_{+}$,
with complex numbers $c^{(\cdot)}_{\pm}$ and two additional boundary
fermions $a_{0}^{}$ and $a_{N+1}^{}$. We note that the boundaries
considered here act as pure back-scatterers \cite{ref1,ref2}. Thus a
combination with the forward-scattering impurities of
Refs.~\cite{ref8,ref9} may result in physically relevant, yet
completely integrable impurities.  Corresponding Hamiltonians may be
constructed by the methods outlined here \cite{ref10}.
    \vspace{0.6cm}\newline{\small 
{\bf Acknowledgments}:~~R.A.R. gratefully acknowledges financial support by the DFG through
Sonderforschungsbereich 393; G.X.W. would like to thank
Institut f\" ur Physik, Technische Universit\"{a}t Chemnitz
for kind hospitality.
    }

    \end{document}